# Life Estimation of HVDC Cable Insulation under Load Cycles: from Macroscopic to Microscopic Charge Conduction Modelling

Bassel Diban, *Member, IEEE*, and Giovanni Mazzanti *Fellow, IEEE*

*Abstract*—This paper goes one step forward in the life estimation of HVDC cable insulation under load cycles by introducing for the first time a microscopic model of charge conduction and transport – i.e., Bipolar Charge Transport BCT model – for electric field calculation inside the insulation thickness. The paper firstly includes the development and the validation of BCT model with that found in literature. Then, the parameters of the developed BCT model are optimized using Pulsed Electro-Acoustic (PEA) space charge measurements. Followed by the integration of the developed, validated and optimized model into the electric field calculation for life estimation of a 500 kV DC-XLPE insulated cable subjected to Type Test load cycles according to Cigrè Techical Brochure 852. The developed microscopic model is compared to the macroscopic models already found in the literature. The microscopic model shows a comparable electric field inversion similarly to macroscopic models. However, the behavior of the microscopic model is noticed to be different under heating and cooling load cycles. In hot cable, the maximum electric field stabilizes at different amplitude and position inside the insulation thickness in both models. This investigation has been carried out in the framework of the HEU-NEWGEN research project.

## I. INTRODUCTION

THE reliability of High Voltage Direct-Current HVDC cables has received increasing attention during recent decades as a result of the widespread adoption of HVDC cable systems as well as the expansion of renewables deployment worldwide [1]. Life and reliability estimation has been developed to predict the life loss in cable insulation after being subjected to various stresses i.e., electrical, thermal, mechanical, and environmental [2]. Electrothermal life models are mostly used in literature both in cable design and in the estimation of remaining life [2],[3],[4]. Those models and their relevant parameters (e.g., Voltage Endurance Coefficient VEC) can be verified experimentally using Accelerated Life Tests ALTs performed at high electric fields [5], whose results need to be carefully processed by ad hoc statistical tools under the Weibull hypothesis for breakdown strength and failure time distributions [6]. Although the empirical electrothermal life models are considered the most direct approach of life estimation, they are highly dependent on variables that are difficult to calculate straightforwardly i.e., the temperature and the electric field distributions inside the insulation thickness. Two methods of electric field calculation were introduced in literature, i.e., macroscopic and microscopic models of charge conduction. The macroscopic models are described by an analytical closed-form dependence of the electrical conductivity on temperature and electric field [7], while the microscopic models physically describe the charge conduction processes, i.e., injection, mobility [8], trapping, de-trapping, and recombination as in the Bipolar Charge Transport BCT model [9] whose parameters are usually derived using space charge measurements and/or conductivity measurement over time [10],[11]. So far, authors have only used the macroscopic models of electrical conductivity for electric field calculation dedicated to life and reliability estimation of HVDC cables [2],[4],[12],[13]. This paper employs, for the first time, microscopic models of charge conduction and transport in the procedure for life and reliability estimation of HVDC cables under load cycles. The paper is organized as follows:

- Section II, the microscopic BCT model is developed as in the literature [9],[14].
- Section III, the microscopic BCT model and the resulting electric field are validated using a replicated case study as in [9],[14].
- Section IV, the BCT model parameters are found using an ad-hoc MATLAB optimization process with the space charge measurements carried out on DC-XLPE (Cross-linked Polyethylene dedicated for DC cables).
- Section V, the developed and optimized BCT model is included in the electro-thermal life and reliability estimation procedure to calculate the electric field.

Corresponding author: Bassel Diban.
Bassel Diban and Giovanni Mazzanti are with the University of Bologna, Bologna, BO 40136 Italy (e-mail: bassel.diban2@unibo.it giovanni.mazzanti@unibo.it ).
Color versions of one or more of the figures in this article are available online at http://ieeexplore.ieee.org

- Section VI includes an in-depth discussion of the main similarities and differences between both microscopic and macroscopic models as well as the resulting electric field distributions.
- Section VII is a conclusion that highlights the main findings and future research.
- Appendix includes details about the space charge measurements carried out on DC-XLPE together with the pattern of the optimized BCT model at each temperature, electric field and voltage polarity.

This investigation has been performed in the framework of the WP4 "Tools and models for reliable and resilient HVDC cable systems" of the HEU-NEWGEN research project.

## II. BCT Model Development

The BCT model includes the following physical processes [9],[11],[14]:
- the injection of charge carriers from the electrodes into the insulation, represented here by Schottky injection of both electrons and holes, as in (1);
- the mobility of charge carriers from the vicinity of the injecting electrode towards the opposite electrode (2), (3);
- the various physical processes that lead to a variation in the charge density, i.e., trapping, detrapping, and recombination of charge carriers (4) – (12).

$$J_{sch\ e,h} = A_0 T^2 exp\left(-\frac{w_{e,h\ i}}{k_B T}\right)\left[exp\left(-\frac{e}{k_B T}\sqrt{\frac{eE}{4\pi\varepsilon}}\right) - f_s\right] \quad (1)$$

$$\mu_{e,h} = \frac{2v\ a}{E} exp\left(-\frac{w_{\mu_{e,h}}}{k_B T}\right) \cdot sinh\left(\frac{q\ a\ E}{2k_B T}\right) \quad (2)$$

$$v = \frac{k_B T}{h} \quad (3)$$

Source terms are given by the following equation:

$$s_{e,\mu} = -B_e \rho_{e,\mu}\left(1 - \frac{\rho_{e,t}}{\rho_{e0t}}\right) + D_e \rho_{e,t} - S_1 \rho_{e,\mu}\rho_{h,t} - S_3 \rho_{e,\mu}\rho_{h,\mu} \quad (4)$$

$$s_{e,t} = B_e \rho_{e,\mu}\left(1 - \frac{\rho_{e,t}}{\rho_{e0t}}\right) - D_e \rho_{e,t} - S_0 \rho_{e,t}\rho_{h,t} - S_2 \rho_{e,t}\rho_{h,\mu} \quad (5)$$

$$s_{h,\mu} = -B_h \rho_{h,\mu}\left(1 - \frac{\rho_{h,t}}{\rho_{h0t}}\right) + D_h \rho_{h,t} - S_2 \rho_{h,\mu}\rho_{e,t} - S_3 \rho_{e,\mu}\rho_{h,\mu} \quad (6)$$

$$s_{h,t} = B_h \rho_{h,\mu}\left(1 - \frac{\rho_{h,t}}{\rho_{h0t}}\right) - D_h \rho_{h,t} - S_0 \rho_{h,t}\rho_{e,t} - S_1 \rho_{h,t}\rho_{e,\mu} \quad (7)$$

Where:

$$S_0 = S_{0,base} \quad (8)$$

$$S_1 = S_{1,base} + \frac{\mu_e}{\varepsilon} \quad (9)$$

$$S_2 = S_{2,base} + \frac{\mu_h}{\varepsilon} \quad (10)$$

$$S_3 = S_{3,base} + \frac{\mu_e + \mu_h}{\varepsilon} \quad (11)$$

$$D_{e,h} = v\ exp\left(-\frac{w_{tr\ e,h}}{k_B T}\right) \quad (12)$$

Current continuity equations are derived from the following:

$$\nabla \cdot (J_{drift} + J_{diff}) + \frac{\partial \rho_{e,h}}{\partial t} = s_{e,h\ \mu,t} \quad (13)$$

Drift current density is written as follows:

$$J_{drift\ e,h\ \mu}(r,t) = \mu_{e,h}(r,t)\ \rho_{e,h\ \mu}(r,t)\ E(r,t) \quad (14)$$

Diffusion current density is given by:

$$J_{diff\ e,h\ \mu}(r,t) = D_{diff} \nabla(\rho_{e,h\ \mu}) \quad (15)$$

$$D_{diff} = \frac{k_B T}{q}\mu_{e,h} \quad (16)$$

The net charge density is given as follows:

$$\rho_{net} = -\rho_{e,\mu} + \rho_{h,\mu} - \rho_{e,t} + \rho_{h,t} \quad (17)$$

The electric field is calculated from:

$$\nabla \cdot (\varepsilon_0 \varepsilon_r E) = \rho_{net} \quad (18)$$

Where: $S_0, S_1, S_2, S_3$ are the recombination coefficients, $S_{0,base}$ is the base level for trapped hole-trapped electron recombination, which exhibits no velocity/field dependency, as both carriers are stationary. $S_{1,base}$, $S_{2,base}$ and $S_{3,base}$ are base level recombination rates for trapped hole–mobile electron, mobile hole–trapped electron and mobile electron–mobile hole recombination, respectively. $v$ attempt to escape frequency, $h$ Plank's constant. $B_e$ and $B_h$ are trapping coefficients for electrons and holes, respectively. $\mu_{e,h}$ is the mobility of charges. $w_{\mu_a}$ is the depth of a single trapping level for mobility (eV). $a$ is the average distance between traps. $D_{e,h}$ are the de-trapping coefficients for electrons and holes, respectively. $\rho_{e0t}$ and $\rho_{h0t}$ the maximum trapped charge densities for electrons and holes, respectively. $J$ is the conduction current density (A/m$^2$). The subscripts $\mu$, $t$ refer to mobile and trapped charges, respectively. $\rho$ refers to the charge density (C/m$^3$), $\rho_{net}$ is the net charge density at a generic point and time instant (C/m$^3$), $E$ is the electric field (kV/mm).

## III. BCT Model Validation

BCT model – developed in this study and used for electric field calculations and life estimation under DC conditions – has been validated using the results of similar simulations found in the literature by S. Le Roy [9], G. Chen [14]. A 4.5-mm-thick 90-kV XLPE cable insulation is used for model validation. This validation used Finite Difference Method (FDM) with 100 mesh points inside the cable insulation for BCT and electric field calculations. Temperature drop inside the insulation is generated by the thermal ohm's law between $T(r_i) = 65°C$ and $T(r_o) = 45°C$, respectively.

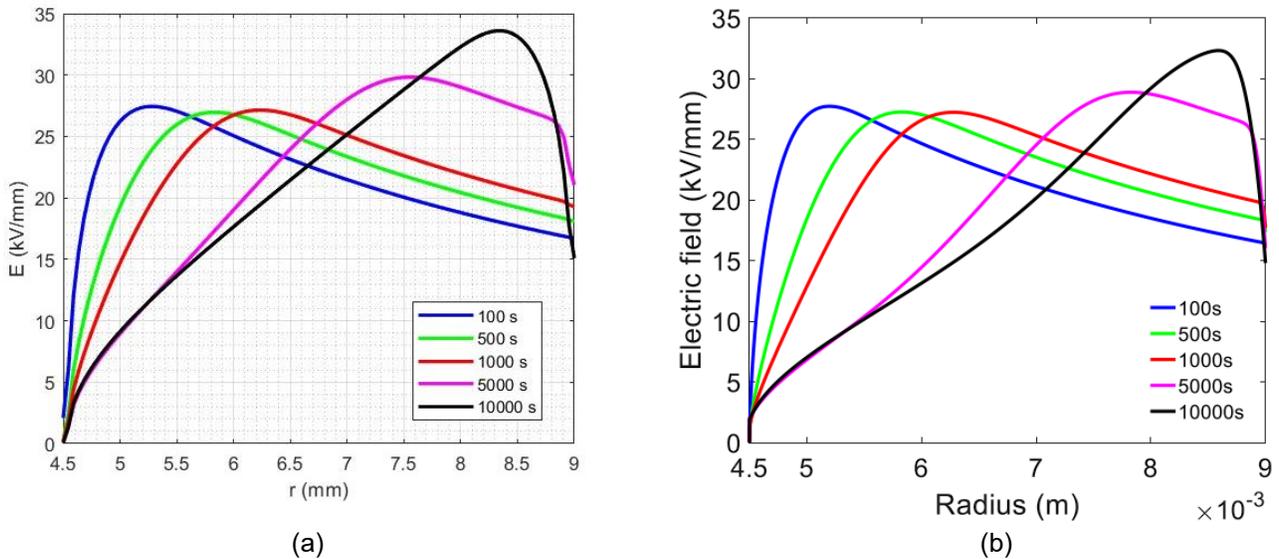

Fig. 1. The electric field distribution for the validation of the BCT model developed in this work (a) with that developed in S. Le Roy et. al. [9] and G. Chen et. al. [14] (b), having the same temperatures, cable geometry and applied voltage.

$$T(r) = T(r_i) - [T(r_i) - T(r_o)] \ln\left(\frac{r}{r_i}\right) / \ln\left(\frac{r_o}{r_i}\right) \quad (19)$$

Five time points are considered for plotting the electric field, i.e., 100 s, 500 s 1000 s, 5000 s, and 10000 s, for the sake of comparison. Fig. 1 shows the comparison between the electric fields calculated from the BCT model developed in this study (in Fig. 1(a)) with that calculated from the model developed in [9], [14], (in Fig. 1(b)). The electric field distributions are in quite good agreement with a maximum electric field of 33.5 kV in this model compared to 32.7 kV reported in the literature. Table I reports a numerical comparison between the peaks of the electric field distributions (since they are critical in the life estimation) at different times along with the relevant relative errors. The difference between the two models increases from 1.25% at the beginning of the simulation to arrive to 2.5% at t=5000 s. Those variations might be justified by the absence of reporting some parameters in [9] and [14] (e.g., distance between traps in the charge mobility equation). However, the values obtained in this study at t=5000 s and at t=10000 s are slightly higher, thus they can be deemed as conservative.

TABLE I
VALIDATION OF BCT MODEL WITH LITERATURE

| Time (s) | $E_{max}$ (kV/mm) | | Relative error (%) |
|---|---|---|---|
| | This model | Literature [9],[14] | |
| 100 | 27.45 | 27.80 | 1.26 |
| 500 | 26.96 | 27.30 | 1.25 |
| 1000 | 27.15 | 27.30 | 0.55 |
| 5000 | 29.83 | 29.10 | 2.51 |
| 10000 | 33.50 | 32.70 | 2.45 |

## IV. BCT MODEL PARAMETERS OPTIMIZATION

The parameters of the microscopic model are investigated by iterative fitting of MATLAB simulations in comparison with the charge density measured using the PEA (Pulsed Electro-acoustic) space charge measurement carried out on DC-XLPE. 200-μm-thick hot-pressed DC-XLPE specimens were produced by VTT research center of Finland, coordinator and partner of the HEU-NEWGEN research project. Then, the specimens were degassed in a vacuum oven at 70°C for 48 h to remove the cross-linking by-products. The PEA space charge measurements were carried out at the University of Bologna, Italy (leader of WP4 of the HEU-NEWGEN research project) under different temperature, electric fields, and polarity of the applied voltage. However, the space charge measurement during volts-on period shows – in addition to the trapped/mobile space charges inside the specimen – induced charges on the electrode that cannot be represented by BCT model. Therefore, capacitive and image surface charges are superimposed on the simulation results to be comparable with the experimental data [10],[15].

$$n_{cap} = \pm \varepsilon E_{mean} \quad (20)$$

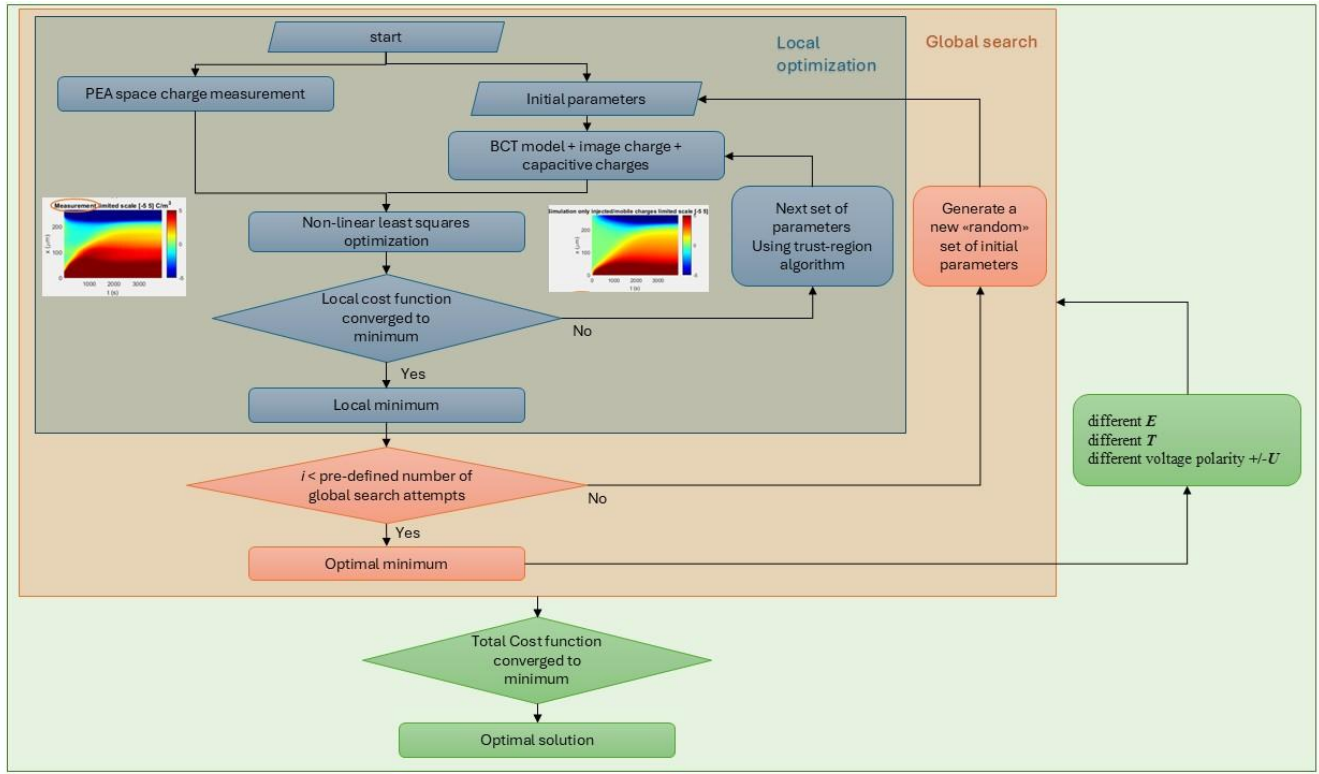

Fig. 2. Algorithm for local and optimal (best-known) minimum search for the parameters of bipolar charge transport model.

$$n_{im}(-) = -\frac{1}{x_d} \sum_{i=1}^{i=d-1} \rho_{x_i} \cdot (x_d - x_i) \cdot \Delta x_i \quad (21)$$

$$n_{im}(+) = -\frac{1}{x_d} \sum_{i=1}^{i=d-1} \rho_{x_i} \cdot (x_i) \cdot \Delta x_i \quad (22)$$

Where $n_{cap}$ is the capacitive contribution to surface charges (C/m$^2$), $n_{im}$ is the image surface charges (C/m$^2$), $x_i$ is the distance at the point $i$ from the cathode. $x_d$ is the insulation thickness divided into $d-1$ intervals. $\Delta x_i$ is the interval distance. $\rho_{x_i}$ is the volume charge density (C/m$^3$).

Figure 2 shows a flowchart that illustrates the optimization algorithm. The optimization depends on the non-linear least squares optimization which searches the parameters that lead to the minimum of the sum of the squared differences between the space charge measurement and the simulation. The minimization of the cost function is achieved using trust-region algorithm [10]. MATLAB$^{TM}$ is used for both the simulation and the optimization processes. An ad-hoc Graphic User Interface (GUI) has been developed to follow the evolution of the optimization process at each iteration (as in Figure 3).

The insulation thickness in the simulation is divided into 50 points, similarly, the measured data is interpolated to give 50 points inside the insulation thickness. The duration of the measurement considered for the optimization is ≈ 4000 seconds as a compromise between the stability of the pattern of the space charge measurement and the duration of each iteration during the optimization process of the simulation parameters. In the optimization code, BCT model discretization is modified to consider for flat geometry to be compared with the space charge pattern in flat specimens. The local minimum is the first minimum at which the algorithm arrives starting from the initial parameters. The optimal solution is the best-known minimum as a result of the repetition of the same algorithm starting from a pre-defined range of random set of initial parameters.

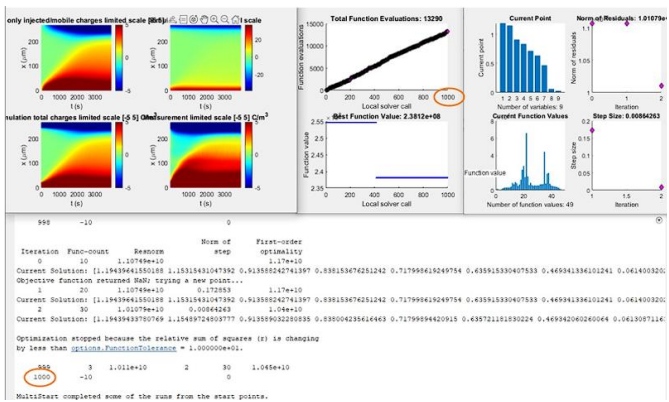

Fig. 3. Optimization GUI window evolution during the optimization process at E=+40 kV/mm and T=20°C for DC-XLPE from HEU-NEWGEN Project.

Figure 3 shows the GUI evolution during the optimization process at E=+40 kV/mm and T=20°C for DC-XLPE specimens from HEU-NEWGEN project. For more details about the optimization results, Figures A1–A4, in the Appendix, show the patterns of the PEA measurement (in Figures A1(a)–A4(a)) in comparison with the optimized parameters (in Figures A1(b)–A4(b)) at different temperatures, electric fields, polarity of the applied voltage.

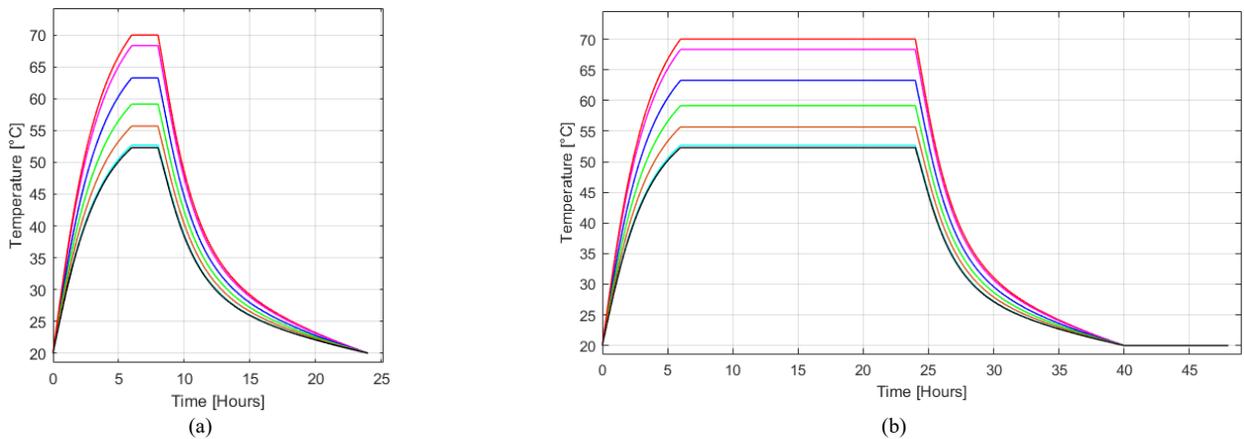

Fig. 4. Temperature profile during the 24-hour (a) and the 48-hour (b) Type Test load cycles for 5 radial points inside the insulation thickness of the case-study cable according to [16].

Table II shows the optimization result of the parameters of BCT model for DC-XLPE specimens from HEU-NEWGEN project as a function of electric field $E$ (ranging from -30 kV/mm) and temperature T (ranging from 20°C to 50°C). Higher temperatures (including the design temperature of the state-of-the-art HVDC cables) result in a very fast charge dynamic pattern that cannot be recognized by the optimization algorithm. It can be noticed from Table II that the optimized parameters show comparable injection and mobility (energy barriers) for both electrons and holes, compared to the parameters found in the literature, in the sense that the optimized parameters exhibit fairly similar values for electrons and holes while the literature parameters are more different for electrons and holes. The latter might be justified by the inclusion of the negative polarity of the applied voltage in the optimization which reduces the geometrical and material effect of the PEA test cell electrodes. The optimized parameters will be further considered in the electric field calculation for the life estimation in the following sections.

TABLE II
OPTIMIZED PARAMETERS OF BCT MODEL FOR DC-XLPE FROM HEU-NEWGEN PROJECT

|  | $E$ (kV/mm) | $T$ (°C) | $w_{ei}$ (eV) | $w_{hi}$ (eV) | $w_{tre}$ (eV) | $w_{trh}$ (eV) | $w_{ue}$ (eV) | $w_{uh}$ (eV) | $B_e$ (1/s) | $B_h$ (1/s) | $S_{base}$ (m$^3$/s.C) |
|---|---|---|---|---|---|---|---|---|---|---|---|
| Initial | -- | -- | 1.18 | 1.19 | 0.84 | 0.91 | 0.671 | 0.652 | 0.29 | 0.16 | 0.047 |
| Local min | -30 | 20 | 1.18 | 1.18 | 0.84 | 0.91 | 0.672 | 0.652 | 0.30 | 0.16 | 0.046 |
|  | +30 |  | 1.20 | 1.18 | 0.91 | 0.85 | 0.659 | 0.653 | 0.17 | 0.30 | 0.046 |
|  | +40 |  | 1.20 | 1.14 | 0.92 | 0.84 | 0.654 | 0.686 | 0.15 | 0.29 | 0.042 |
|  |  | 50 | 1.22 | 1.22 | 0.90 | 0.89 | 0.687 | 0.682 | 0.40 | 0.49 | 0.048 |
| Optimal solution | Multi E, T and polarities |  | 1.22 | 1.20 | 0.91 | 0.90 | 0.684 | 0.680 | 0.30 | 0.30 | 0.045 |
| Literature values [9] |  |  | 1.27 | 1.16 | 0.96 | 0.99 | 0.71 | 0.65 | 0.1 | 0.2 | - |
| Literature values [10] |  |  | 1.21 | 1.1 | - | - | 0.71 | 0.6 | 0.2 | 0.9 | - |
| Literature values [11] |  |  | 0.905 | 1.148 | 1.03 | 1.03 | 0.76 | 0.74 | - | - | 0.0001 |

## V. CASE STUDY FOR LIFE ESTIMATION

### A. Cable Characteristics:

The case-study investigates the electro-thermal life of a 500kV-HVDC extruded cable subject to the Type Test (TT) load cycles according to [16]. The main electrical and thermal characteristics of the case-study cable, as well as the parameters of the electro-thermal life model used for their DC-XLPE insulation, can be found in Table III.

TABLE III
CASE-STUDY CABLE CHARACTERISTICS

| Parameter | value |
|---|---|
| Rated power (monopolar scheme) (MW) | 960 |
| Rated current, $I_n$ (A) | 1920 |
| Conductor Material | Cu |
| Conductor cross-section (mm$^2$) | 2000 |
| Rated voltage, $U_0$ (kV) | 500 |
| Conductor design temperature $T_D$ (°C) | 70 |
| Insulation Material | DC-XLPE |
| relative permittivity $\varepsilon_r$ | 2.3 |
| Inner semiconductor thickness (mm) | 2 |
| Insulation thickness (mm) | 28.1 |
| Outer semiconductor thickness (mm) | 1 |
| Metallic shield thickness (mm) | 1 |
| Thermoplastic sheath thickness (mm) | 4.5 |
| Burial depth $bb$ (m) | 1.3 |
| Design value of soil resistivity (m.K/W) | 1.3 |
| Ambient temperature (°C) | 20 |
| Design life $L_D$ (years) | 40 |
| Design failure probability $P_D$ (reliability $R_D$) % | 1 (0.99) |
| $n_D$ IPM exponent at design temperature | 10 |
| $B$ Arrhenius model parameter (K) | 12430 |
| $b_{ET}$ electro-thermal synergism parameter (K) | 0 |

### B. Temperature Profile

The cable is subjected to the Type Test (TT) load cycles according to CIGRÉ Technical Brochure 852 [16]. The TT load cycles consist of 24 consecutive cycles of the 24-hour type, followed by 3 cycles of the 48-hour type. Figure 4 shows the temperature profile in 5 points inside the insulation thickness during the 24-hour load cycle (in Figure 4(a)) and 48-hour load cycle (in Figure 4(b)).

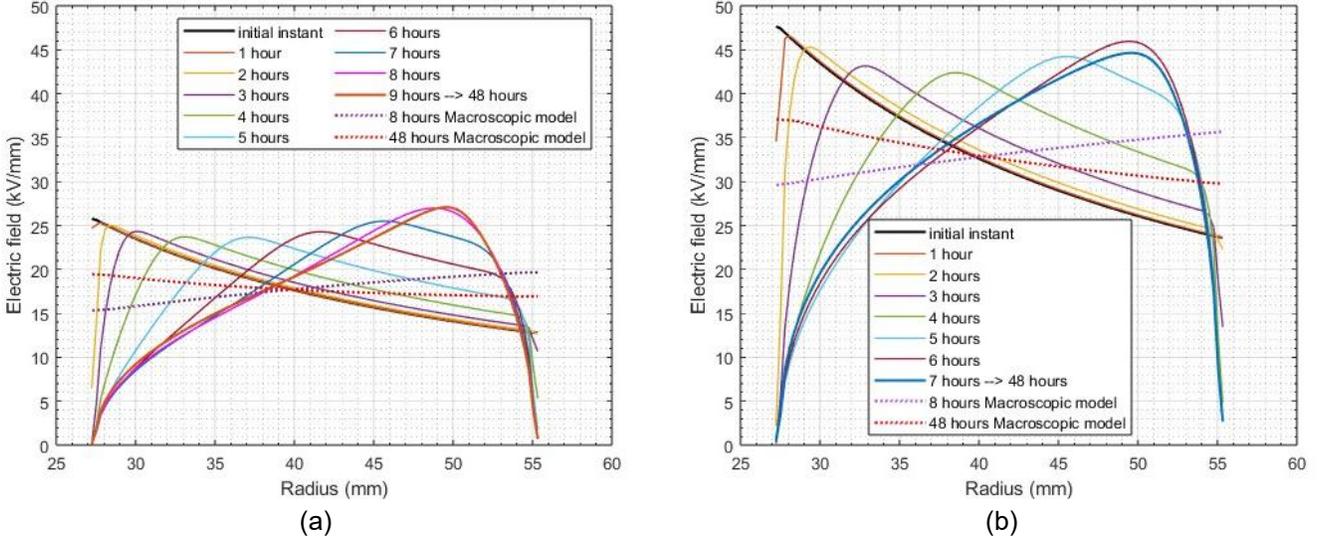

Fig. 5. Transient electric field distribution inside the insulation thickness of the case-study cable during 48-hour load cycle under (a) rated voltage $U_0$ (b) the Type Test voltage $U_{TT}$

### C. Electric field calculation:

The transient electric field profiles $E(r,t)$ are calculated during one load cycle of the LC period at 100 points inside the insulation of the studied cable during both 24-hour and 48-hour load cycles LCs. Two methods are used for the calculation:
1) the microscopic method consisting in solving BCT Equations (1) – (18) for the BCT parameters obtained during NEWGEN project and reported in Table II (solid lines in Fig. 5);
2) the macroscopic method based on the solution of Maxwell's equations with Klein's closed-form dependence of the electrical conductivity on temperature and electric field (dotted lines in Fig. 5).

With the microscopic method Figure 5 shows that the electric field profile within cable insulation starts from the Laplacian field at the beginning of the cycle, then it takes approximately 9 hours and 7 hours at $U_0$ and $U_{TT}$, respectively, to arrive at a steady-state field distribution, which is then kept until the end of the 48-h load cycle. This result is different compared to that obtained with the macroscopic method, where the electric field profile within cable insulation during the 48-h load cycle changes from the Laplacian at the beginning to the "hot cable" field distribution in the highest temperature part of the cycle (see the dashed line at 8 h, where the field profile is completely inverted) and then back to the "cold cable" field distribution in the lowest temperature part of the cycle (see the dashed line at 48 h, where the field profile is not inverted).

The most electrically stressed point inside the insulation starts at the inner insulation at the beginning of the LC because of the Laplacian electric field in the cylindrical cable geometry, then over time, the most stressed point shifts towards the outer insulation until it arrives to ≈70% of the entire insulation thickness at 9 hours and 7 hours at $U_0$ and $U_{TT}$, respectively.

Thereafter, the most stressed point remains there throughout the cycle. Also this result is different from that obtained with the macroscopic method, where the most stressed point – consistently with the whole electric field profile – during 48-h load cycles changes from the outer insulation surface for the "hot cable" in the highest temperature part of the cycle (dashed line at 8 h) to the inner insulation surface for the "cold cable" in the lowest temperature part of the cycle (dashed line at 48 h).

The longer time to steady state field obtained at $U_0$ (9 h) compared to $U_{TT}$ (7 h) is easily explained via the macroscopic conductivity approach, since the value of dielectric constant $\tau = \varepsilon/\sigma(E,T)$ is expected to be greater for the lower voltage $U_0$ compared to $U_{TT}$: indeed, at a given temperature, the greater the voltage and the field $E$, the greater the conductivity $\sigma(E,T)$ and the smaller the dielectric constant $\tau$.

### D. Life Estimation:

The electrothermal IPM-Arrhenius life model of eq. (23) is used to assess the lifespan of the DC-XLPE cable insulation under TT conditions. Specifically, cast into the Miner's law framework of aging accumulation over time [13] as per the following eqns. (24)-(26), it estimates the accumulated loss of life $LF_{cycle}(r)$ and the time-to-failure $L_{cycle}(r)$ vs. insulation radius $r$ during TT load cycles, as well as the life of the most stressed point in the insulation $L_{cycle}$:

$$L(E,T) = L_D \cdot [E/E_D]^{-(n_D - b_{ET}T'')} [E_D/E_0]^{b_{ET}T''} e^{-BT''} \quad (23)$$

$$LF_{cycle}(r) = \int_0^{t_d} \frac{dt}{L[E(r,t),T(r,t)]} = \frac{1}{K_{cycle}(r)} \quad (24)$$

$$L_{cycle}(r) = t_d \times K_{cycle}(r) \quad (25)$$

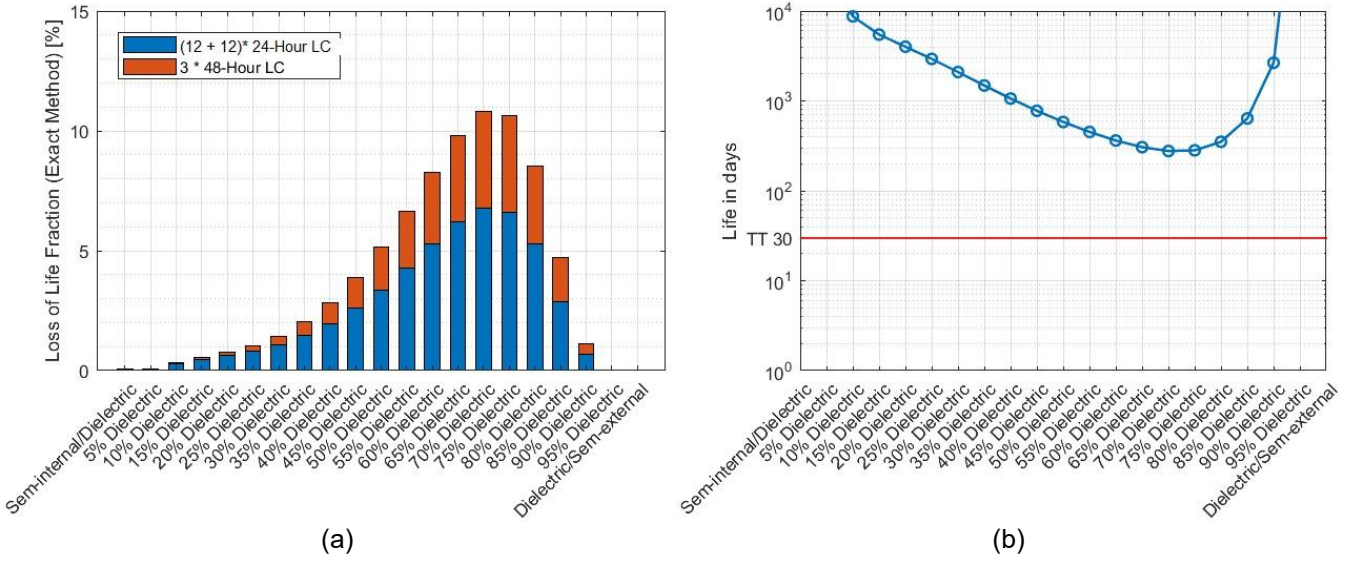

(a)            (b)

Fig. 6. (a) Percent loss of life fractions inside the insulation thickness of the case-study cable under the Type Test conditions during all 24-hour load cycles (blue bars) and 48-hour load cycles (red bars), (b) Life distribution (in days) inside the insulation thickness of the case-study cable under the Type Test conditions.

$$L_{cycle} = min\{L_{cycle}(r), r \in [r_i, r_o]\} \quad (26)$$

where: $L$ is life at DC electric field $E$ and temperature $T$, $T''=1/T_D-1/T$ ($E_D$, $T_D$ and $L_D$ being design electric field, temperature and life respectively), $n_D$ is the value of voltage endurance coefficient (VEC) at $T_D$, $b_{ET}$ is the synergism factor between the electrical and thermal stress, $B = \Delta W/K_B$, $\Delta W$ is the activation energy of the main thermal degradation reaction, $K_B = 1.38 \times 10^{-23}\ J/K$ is the Boltzmann constant, $K_{cycle}(r)$ is the number of cycles-to-failure at a generic radius $r$. Further details about the detailed temperature calculations, macroscopic electric field calculation, life and reliability estimation are omitted here for the sake of brevity, and can be found in [2], [4],[12].

Figure 6(a) shows the loss of life distribution inside the insulation thickness of the case-study cable during 24-hour LCs in blue bars and 48-hour LCs in red bars, obtained using the BCT model. While Figure 6(b) shows the life of cable (in days) inside the insulation thickness of the case-study cable under the Type Test conditions. Consistently with the field profiles reported in Figures 5(a) and 5(b), the most stressed point – i.e. the one with the highest loss of life - corresponds to 70% of the insulation thickness at a maximum loss of life = 13 %. A quite significant loss of life is associated with the 3 48-h load cycles, compared to the 24 24-h load cycles.

In Figure 6(b), the minimum life point, corresponding to 70% of the insulation thickness, represents the life of cable according to BCT model, and still greater than the duration of the TT. Therefore, the cable is expected to withstand TT according to the life estimation using the BCT model. The same was concluded in [13] for 500 kV cable using the macroscopic conductivity approach, except the position of the maximum stress point inside the insulation, where it is always placed either at the inner insulation or at the outer insulation in the case of macroscopic models – as hinted at above.

## VI. DISCUSSION

DC electrical conductivity is considered one of the most critical parameters in determining the performance of a certain insulating material in DC systems. The main reason is the dependency of the electric field distribution on the conductivity in DC cable systems as the poling DC field causes space charge mobility, trapping and/or de-trapping inside the cable insulation, hence changing the capacitive electric field distribution into a more complicated field distribution, dissimilar to the AC system where the electric field is still capacitive due to the absence of such space charge accumulation and mobility. Physical models for describing the space charge behavior inside the insulation are developed with the proper fitting parameters to the space charge measurements (e.g., BCT model).

Another simpler method for describing the space charge behavior is to be represented by the electrical conductivity in a phenomenological way, where the electrical conductivity is expressed by means of a closed-form function of the temperature and electric field. Then the charge density can be calculated as the charges accumulated at the discontinuities of permittivity and conductivity.

The former methods are called physical (or microscopic) models, while the latter are called phenomenological (or macroscopic) models for DC electrical conductivity. Macroscopic and microscopic models of charge conduction are characterized in this paper as reported in Table IV:

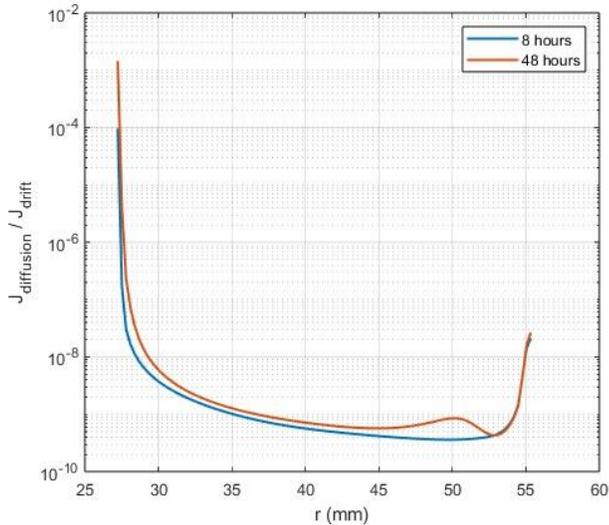

Fig. 7. Ratio between diffusion current density / drift current density inside the insulation for hot and cold cables during the load cycle.

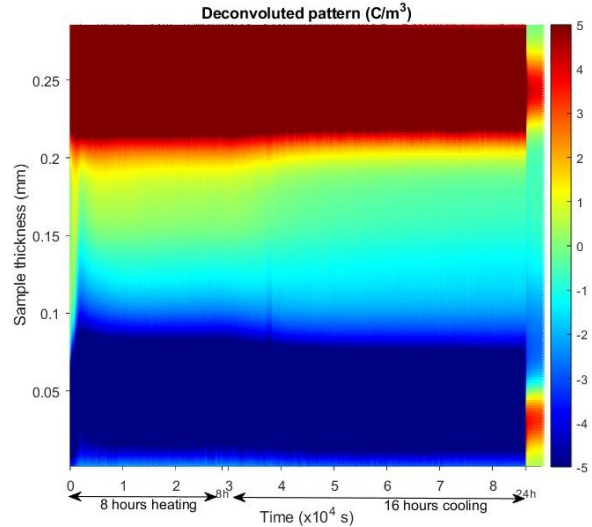

Fig. 8. Space charge measurement on DC-XLPE flat specimen subjected to 24-hour load cycle (8 hours heating followed by 16 hours cooling).

TABLE IV
CHARACTERISTICS OF MACROSCOPIC AND MICROSCOPIC CHARGE CONDUCTION MODELS

|  | Macroscopic | Microscopic |
|---|---|---|
| Source term | $\nabla \cdot J + \frac{\partial \rho}{\partial t} = 0$ | $\nabla \cdot J_{e,h} + \frac{\partial \rho_{e,h}}{\partial t} = s_{e,h}$ |
| Conductivity dependency | $f(E,T)$ | $f(\mu_{e,h}, \rho_{e,h}\mu)$ |
| Space charge distribution | Uniform | Non-uniform |
| Memory of generated space charges | Memory-less | Has a memory |
| Electric field inversion | Present and moderate | Present and strong |
| Electric field response to heating-cooling load cycles | Significant | Minor or no change |
| Maximum electric field position | Either at inner or outer insulation | Can be inside the insulation |

Figure 5 shows that, according to the BCT, the cooling period has no effect on the electric field distribution, dissimilar to the electric field derived from macroscopic models which varies according to temperature variation during the load cycle. This response can be justified by the memory dynamic effect of the microscopic models that depend on the history of space charges variation, compared to the memory-less macroscopic models that depends only on the electric field and temperature, as already illustrated in Table IV.

It is worth noting that considering charge diffusion with various diffusion parameters shows no variation of the electric field during the cooling period of the load cycle as long as the voltage is still applied. Charge diffusion has been deemed negligible in [17] and [10], while it shows noticeable charge relaxation in [11], however, in a thin specimen over many hours and when voltage is off. Figure 7 shows the diffusion current density / drift current density ratio inside the insulation thickness for hot and cold cables (at 8 and 48 hours during the load cycle, respectively). It shows that diffusion current density is lower than drift current density by many orders of magnitude, i.e., ranging from 3 to 9 orders of magnitude. The drift conduction dominates the electric field distribution, although the diffusion is present in the high charge density gradient zones.

As an extension to this discussion, a long space charge measurement on 0.22-mm-thick DC-XLPE flat specimen subjected to 35 kV/mm and one 24-hour load cycle (8 hours heating followed by 16 hours cooling) is carried out (as in Figure 8). It shows that space charge variation during the cooling period is minor and mainly affects the electrodes vicinity. Future simulative and experimental investigations are necessary to verify this finding for any possible charge relaxation phenomena that might affect the electric field distribution during the load cycle. This includes considering other BCT model developments e.g., a distributed energy level of traps, as in [18], or even future developments – for more conservative life estimation – to consider a possible formation of heterocharges as observed experimentally in certain insulating materials under certain conditions. Charge diffusion parameters are also to be investigated experimentally.

## VII. CONCLUSION

This paper studies the electrothermal life estimation in HVDC cables considering the microscopic models of charges conduction for the electric field calculation inside the insulation thickness. The paper first develops and validates the Bipolar Charge Transport (BCT) model using case-studies found in the literature. Then, BCT model is applied for the electric field calculation to be used in the electrothermal life estimation. The novelties in this paper can be summarized as follows:

- The parameters of the BCT model are found using an ad-hoc optimization process in comparison with the experimental data coming from space charge measurements using Pulsed-Electro Acoustic method (PEA) on DC-XLPE flat specimens produced in the framework of the HEU-NEWGEN research project.
- The parameters of BCT optimized in this paper indicate comparable values of both injection and mobility of both holes and electrons, dissimilar to the parameters already found in the literature.
- An in-depth analysis of the electric field distribution using microscopic and macroscopic models under heating and cooling load cycles is introduced for the first time. The main difference is that the electric field inversion is continuous during the consecutive load cycles when using the macroscopic model, while the electric field shows no variation after the first heating of the first load cycle when using the microscopic model. The memory effect of the space charges in the microscopic models can justify this pattern which is extensively discussed in the paper. In hot cable, the maximum electric field point is shifted from the inner insulation to 70% of the insulation thickness using the microscopic model while it is shifted to the outer insulation surface using the macroscopic model.
- The electrothermal life estimation of a case-study HVDC cable under load cycles by means of the electric field calculated via the BCT has been performed for the 1st time here in the literature (using the optimized BCT parameters); similarly to the results obtained with previously-developed macroscopic models, greater life compared to the type test duration is obtained. This agrees with the fact that the well-designed HVDC cables should withstand the type test. The main difference between the macroscopic and microscopic models for the life estimation is the position and the amplitude of the most stressed point inside the insulation thickness. The non-uniformity of the space charge distribution in the case of microscopic models plays an important role in this difference, in comparison with the quasi-uniform charge distribution in the case of macroscopic models.

Future research will focus on the development of the BCT model and the life estimation approach and parameters to consider novel HVDC cable insulations.

## APPENDIX

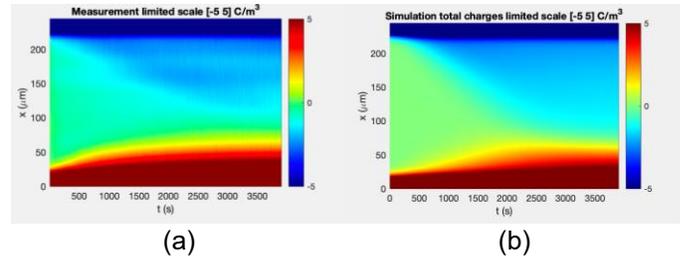

(a)            (b)

Fig. A1. Space charge density (C/m$^3$) in a flat DC-XLPE specimen under negative polarity of the applied voltage (E=-30 kV/mm) and (T=20°C) according to: (a) measurement and (b) optimized simulation from BCT.

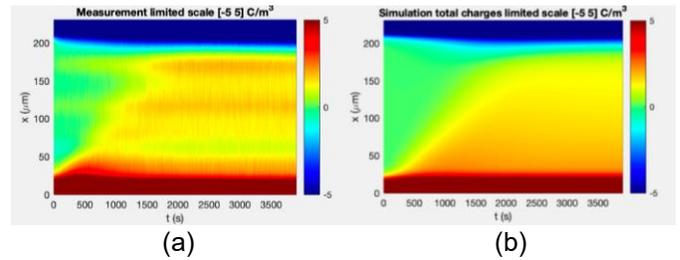

(a)            (b)

Fig. A2. Space charge density (C/m$^3$) in a flat DC-XLPE specimen under positive polarity of the applied voltage (E=+30 kV/mm) and (T=20°C) according to: (a) measurement and (b) optimized simulation from BCT.

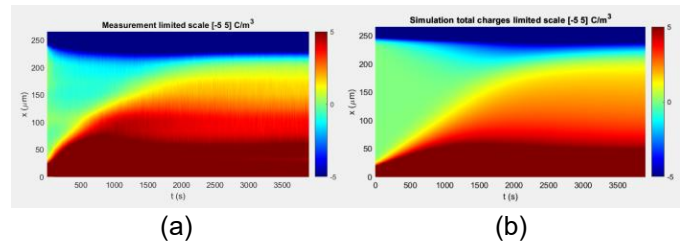

(a)            (b)

Fig. A3. Space charge density (C/m$^3$) in a flat DC-XLPE specimen under positive polarity of the applied voltage (E=+40 kV/mm) and (T=20°C) according to: (a) measurement and (b) optimized simulation from BCT.

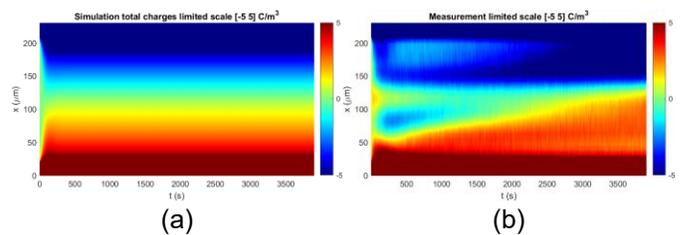

(a)            (b)

Fig. A4. Space charge density (C/m$^3$) in a flat DC-XLPE specimen under positive polarity of the applied voltage (E=+40 kV/mm) and (T=50°C) according to: (a) measurement and (b) optimized simulation from BCT.

## ACKNOWLEDGMENT

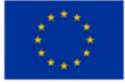

Funded by the European Union Grant Agreement No 101075592. Views and opinions expressed are however those of the author(s) only and do not necessarily reflect those of the European Union or CINEA. Neither the European Union nor the granting authority can be held responsible for them.

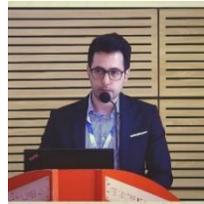

**Bassel Diban** (Member, IEEE) received the MSc and PhD degrees in electrical energy engineering from the University of Bologna, Bologna, Italy (in 2019, and 2023, respectively), where he is currently working as a junior assistant professor. His research interests are life modeling, reliability, diagnostics of HV insulation, and HVDC cable systems.

Dr. Diban is a member of the IEEE DEIS Technical Committee (TC) on "High Voltage Direct-Current (HVDC) cable systems" and CIGRÉ WG B1.91 "Transient Thermal Modelling of Power Cables (update to IEC 60853).

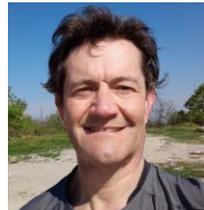

**Giovanni Mazzanti** (Fellow, IEEE) is currently an Associate Professor of HV Engineering and Power Quality at the University of Bologna, Bologna, Italy. He is Consultant to TERNA (the Italian TSO), Rome, Italy. His research interests are reliability and diagnostics of HV insulation, power quality, renewables, and human exposure to EMF. He is author or coauthor of more than 300 papers, and of the book *Extruded Cables for HVDC Transmission: Advances in Research and Development*, (Wiley-IEEE Press, 2013).

Prof. Mazzanti is a member of IEEE PES and DEIS, IEEE DEIS Technical Committee (TC) on "Smart grids," CIGRÉ, CIGRÉ Joint Working Group B4/B1/C4.73 on "Surge and extended overvoltage testing of high voltage direct-current (HVDC) cable systems." He is the Chair of the IEEE DEIS TC on "HVDC cable systems."